\documentclass[aps,pra,reprint,groupedaddress,showpacs]{revtex4-1}

\usepackage{graphicx}
\usepackage{color}
\usepackage{amsmath,amssymb}
\usepackage{bm}
\usepackage{upgreek}
\usepackage{xspace}
\usepackage[colorlinks,urlcolor=blue,citecolor=blue,linkcolor=blue]{hyperref}
\usepackage{tikz}
\usetikzlibrary{matrix,fit}
\tikzset{
  highlight/.style={rectangle,rounded corners,draw,inner sep=0pt,dash pattern=on 3pt off 3pt}
}
\newcommand{\tikzmark}[2]{\tikz[overlay,remember picture,baseline=(#1.base)] \node (#1) {\ensuremath{#2}};}
\newcommand{\submatrixborder}[1][submatrix]{
    \tikz[overlay,remember picture]{
    \node[highlight,fit=(top.north west) (bottom.south east)] (#1) {};}
}

\newcommand{\etal}{et~al.\xspace}
\newcommand{\Rb}{$^{87}$Rb\xspace}
\newcommand{\dphi}{\ensuremath{\Delta \phi}\xspace}

\newcommand{\reffig}[1]{\mbox{Fig.~\ref{#1}}}
\newcommand{\refeq}[1]{\mbox{Eq.~(\ref{#1})}}

\newcommand{\upd}{\text{d}}
\newcommand{\totalD}[2]{\frac{\upd #1}{\upd #2}}
\newcommand{\partialD}[2]{\frac{\partial #1}{\partial #2}}

\newcommand{\abs}[1]{\vert #1 \vert \xspace}

\newcommand{\ket}[1]{\vert #1 \rangle \xspace}

\newcommand{\unit}[1]{\,\mathrm{#1}}
\newcommand{\vect}[1]{\mathbf{#1}\xspace}
\newcommand{\uvect}[1]{\hat{\mathbf{#1}}\xspace}

\begin{document}

\title{Magnetic tensor gradiometry using Ramsey interferometry of spinor condensates}

\author{A.~A.~Wood}
\author{L.~M.~Bennie}
\author{A.~Duong}
\author{M.~Jasperse}
\author{L.~D.~Turner}\email[]{lincoln.turner@monash.edu}
\author{R.~P.~Anderson}

\affiliation{School of Physics and Astronomy, Monash University, Victoria 3800, Australia.}

\date{\today}

\begin{abstract}
We have realized a magnetic tensor gradiometer by interferometrically measuring the relative phase between two spatially separated Bose--Einstein condensates (BECs).
We perform simultaneous Ramsey interferometry of the proximate $^{87}$Rb \mbox{spin-1} condensates in freefall and infer their relative Larmor phase -- and thus the differential magnetic field strength -- with a common-mode phase noise suppression exceeding $50\unit{dB}$.
By appropriately biasing the magnetic field and separating the BECs along orthogonal directions, we measure the magnetic field gradient tensor of ambient and applied magnetic fields with a nominal precision of $0.30\unit{nT\,mm^{-1}}$ and a sensor volume of $2\times10^{-5}\unit{mm}^3$.
We predict a spin-projection noise limited magnetic energy resolution of order $\sim 10\hbar$ for typical Zeeman coherence times of trapped condensates with this scheme, even with the low measurement duty cycle of current BEC experiments.
\end{abstract}


\maketitle

\section{Introduction}
\label{sec:introduction}
Precision measurement of magnetic fields underpins applications as diverse as fundamental symmetry tests~\cite{griffith_improved_2009}, magnetoencephalography~\cite{hari_magnetoencephalography:_2012} and geophysical exploration~\cite{clark_new_2012}.
Many of these applications require precise and accurate measurements of the \emph{change} in magnetic field across a region of space: magnetic gradiometry.
On kilometer scales, magnetic gradiometers remotely sense the field of mineral deposits against the larger but locally homogeneous field of the Earth dipole~\cite{pedersen_gradient_1990}.
On the microscopic scale, optimal magnetic gradiometry of biomagnetic or surface-science sources demands magnetic sensor volumes orders of magnitude smaller.
We present the first microscopic \emph{tensor} magnetic gradiometer based on atomic magnetometry, measuring $\partial B_i/\partial x_j$ (the gradient of vector components along orthogonal axes) with a dynamically configurable baseline.

Tensor measurements incisively probe magnetic source distributions~\cite{zhdanov_3D_2012,clark_new_2012}: the gradient tensor at a single point in space determines the bearing, normalized source-strength, and orientation of a dipole~\cite{pedersen_gradient_1990}.
Tensor gradient magnetometers have to date been macroscopic devices -- employing SQUID~\cite{schmidt_getmag_2004,stolz_magnetic_2006,keenan_high-tc_2010} or fluxgate~\cite{koch_room_1996, sui_compact_2014} sensors -- primarily applicable to geophysics and ordinance detection~\cite{clem_superconducting_1996}.
Magnetic fields also change on much smaller length scales, necessitating equally small magnetic sensors to precisely characterize field variations from microscopic magnetic sources~\cite{stamper-kurn_seeing_2014}.
Spinor condensates are highly sensitive to magnetic field gradients, which typically must be eliminated to observe their rich emergent phenomena: quantum phase transitions, dynamics of topological defects, and fragile macroscopic entangled states~\cite{stamper-kurn_spinor_2013}.
In general, inferring \emph{in vacuo} magnetic field profiles from \emph{ex vacuo} measurements is profoundly difficult, with many applications demanding a direct atomic metric from which to diagnose the cancellation of stray magnetic fields and gradients.
Measuring the magnetic field gradient tensor \emph{in vacuo} provides an almost complete measurement of the local magnetic field landscape of a trapped quantum gas, an indispensable tool for characterizing dephasing mechanisms from inhomogenous magnetic fields.

Precise measurement of magnetic field gradients on small length scales demands small magnetic sensors with high spatiotemporal sensitivity.
Atomic magnetometry is a well-established alternative to SQUIDs and other solid-state magnetometers, delivering absolute, calibration-free measurement of magnetic fields by measurement of the Larmor precession frequency of atomic spins~\cite{budker_optical_2013}.
Warm atomic vapor magnetometers have sensing volumes ranging from hundreds of cubic millimeters for the most sensitive magnetometers down to a few cubic millimeters
achieved with micro-fabricated glass cells~\cite{griffith_femtotesla_2010}.
Elongated clouds of ultracold atoms have been used as time-resolved magnetometers and gradiometers with sub-nT sensitivity and a spatial resolution of $50\unit{\upmu m}$~\cite{koschorreck_high_2011,behbood_real-time_2013}.
Colder, denser clouds of atoms in traps, such as Bose-Einstein condensates (BECs), offer the prospect of precise magnetic measurement on the microscale.
Wildermuth \etal~\cite{wildermuth_boseeinstein_2005} used a highly elongated condensate to measure the magnetically-induced trapping potential variations from a current-carrying wire.
Vengalattore \etal~\cite{vengalattore_high-resolution_2007} used a non-destructive phase contrast imaging technique to spatially resolve Larmor precession in a spinor BEC, attaining sensitivities comparable to SQUID-based devices.
This work establishes tensor gradiometry using highly-sensitive, small-volume atomic magnetometers with a dynamically configurable orientation.

\begin{figure}[t!]
    \centering
    \includegraphics[width=1.00\columnwidth]{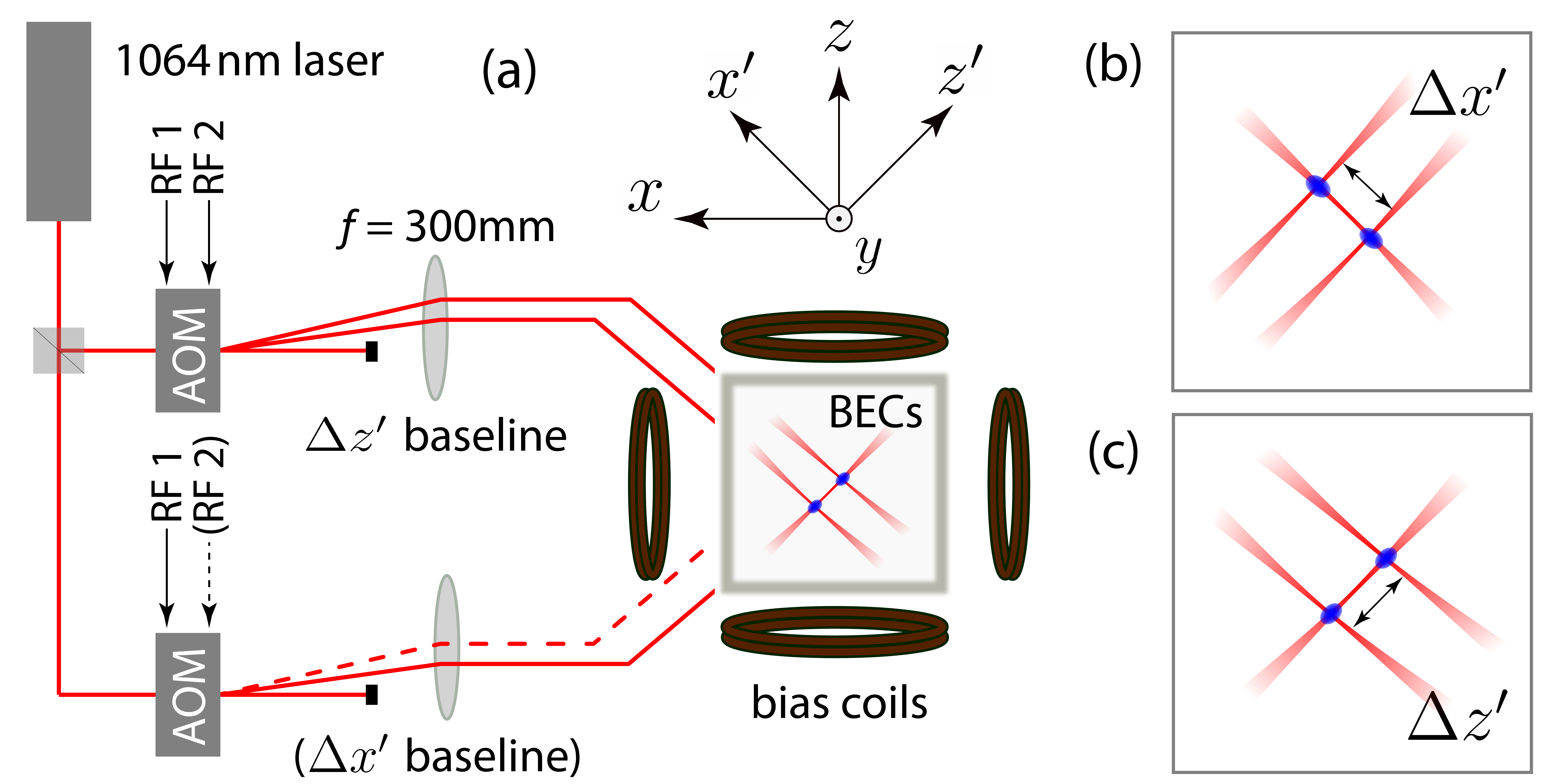}
    \caption{
        \label{fig:apparatus_full}
        (Color online)
        Diagram of the experimental setup (a).
        The gradiometer spans a spatial baseline given by $\Delta x'$ (b) or $\Delta z'$ (c) in the horizontal plane (gravity along $-\uvect{y}$) by translating the pair of condensates in opposite directions along one of the dipole beams, which are oriented at $\sim45^{\circ}$ to the magnetic bias field axes, $x, z$.
        This translation is achieved by feeding a second rf frequency into the AOMs that control the position and amplitude of the dipole beams.
    }
\end{figure}

Our gradiometer measures the phase difference between two spin-1 Ramsey interferometers formed from spatially separated $F = 1$ \Rb BECs.
The use of condensed atoms as magnetic sensors in this case is inessential, though not without benefit: BECs simultaneously offer high atomic density with microscopic sensor volumes, making them ideally suitable as small volume atomic magnetometers.
Each condensate is formed in a separate potential well of a three-beam crossed optical dipole trap and translated to a spatial separation of just less than $1\unit{mm}$ by an acousto-optic modulator (AOM).
We can change our dipole trap configuration so that the gradiometer spans two separation axes in a plane (\reffig{fig:apparatus_full}) and apply bias magnetic fields to make the gradiometer sensitive to field components in all spatial directions.
A Ramsey pulse sequence probes the phase acquired by each condensate over an interrogation time $T$ due to the magnetic field $B$ at the position of each condensate.
The phase difference between fringes from the two interferometers is proportional to the field gradient between the two condensates.
The differential mode of operation of our gradiometer suppresses common-mode noise from transient variations in the magnetic field as well as detection problems.
Simultaneous interrogation of a dual atomic fountain interferometer demonstrated magnetic gradient measurement over a large spatial region~\cite{hu_simultaneous_2011}, with substantial common-mode rejection of noise from drifts and pulse errors.
In this work, we describe the experimental procedure for realizing a tensor gradiometer with a pair of trapped atomic clouds.
We then characterize the gradiometer by measuring the gradient tensor of ambient magnetic fields in our laboratory, as well as measuring the response to applied magnetic field gradients.
Later we describe how the gradiometer configuration can also be used as a prospective microscale co-magnetometer with substantial common-mode rejection.

\section{Differential Ramsey interferometry}
\label{sec:diff_ramsey}
The differential interferometer output $\Delta \phi$ is a measure of the difference in phase acquired by each condensate during the Ramsey sequence, which begins with a $\pi/2$ spin rotation pulse, followed by free evolution over $T$ and closed with a second $\pi/2$-pulse.
The Ramsey sequence simultaneously addresses both condensates, which can be considered as two independent spin-1 interferometers.
For static, spatially varying magnetic fields sampled by two condensates at positions $\vect{r}_1$ and $\vect{r}_2$ we may write
\begin{align}
\label{eq:delta_E_defn_c}
    \frac{\Delta \phi}{ \gamma T} = \vert \vect{B}(\vect{r}_1) \vert - \vert \vect{B}(\vect{r}_2) \vert \approx \nabla \vert \vect{B} \vert_{\vect{r} = \vect{r}_{12}} \cdot (\vect{r}_{1} - \vect{r}_{2}) \, ,
\end{align}
with $\gamma = 2\pi \times 6.996\,\text{kHz}\,\upmu \text{T}^{-1}$ is the gyromagnetic ratio and $\vect{r}_{12} = (\vect{r}_1+\vect{r}_2)/2$.
The differential interferometer is thus sensitive to derivatives of the magnetic field strength $B(\vect{r}) \equiv \vert \vect{B}(\vect{r}) \vert = \sqrt{B_x^2+B_y^2+B_z^2}$,
\begin{align}
\label{eq:field_derivative}
    \frac{\partial B}{\partial x_i} &= \frac{B_x}{B} \frac{\partial B_x}{\partial x_i} +
                                        \frac{B_y}{B} \frac{\partial B_y}{\partial x_i} +
                                        \frac{B_z}{B} \frac{\partial B_z}{\partial x_i} \, .
\end{align}

To achieve vector magnetic field sensitivity along $\vect{x}$ for example, we experimentally null magnetic field components along $\vect{y}$ and $\vect{z}$, leaving a magnetic field with \mbox{$\left|B_x\right| \gg \left|B_y\right|, \left|B_z\right|$}. The measured gradient of $B$ is well approximated by
\begin{align}
\label{eq:modB_approx}
     \frac{\partial B}{\partial x_i} \approx \frac{B_x}{B} \frac{\partial B_x}{\partial x_i}
                                     = \text{sign}(B_x) \partialD{B_x}{x_i} \, .
\end{align}
Two condensates can then be aligned along an axis $x_i$, separated by $\Delta x_i = \vert \vect{r}_1 - \vect{r}_2 \vert$ to measure field differences along that axis. In general we measure components of the magnetic field gradient tensor via the differential Ramsey signal:
\begin{align}
\label{eq:phase_to_gradient}
    \partialD{B_j}{x_i} &\approx \text{sign}(B_j) \frac{1}{\gamma \, \Delta x_i} \totalD{(\dphi)}{T} \, .
\end{align}

The differential interferometer is sensitive to any difference in Zeeman energy between the two condensates.
We perform the interferometry sequence in freefall to prevent spurious contributions to the measured magnetic field gradients from vector light shifts induced by the trapping beams~\cite{happer_effective_1967, romalis_zeeman_1999}.
Imperfect linear polarization of the trapping beams induces an atomic vector polarizability, the spatial variation of which appears as a synthetic magnetic field gradient across each condensate and would lead to dephasing and loss of interferometric contrast.
Additionally, if each condensate experienced a different overall vector light shift due to small intensity differences between the two beams, it would contaminate the gradient measurement.

Freefall interferometry eliminates the vector shift at the expense of introducing a trade-off between the maximum Ramsey interrogation time $T$ and the spatial resolution due to gravity-induced blurring of the sensor volume.
Alternatively, differential Ramsey interferometry of trapped clouds presents a means of precisely measuring and canceling vector light shifts, as discussed in Section~\ref{sec:intrap}.

\section{Experiment}
\label{sec:experiment}
Our experiment begins by forming two \Rb Bose-Einstein condensates in the \mbox{$\ket{F=1, m = -1}$} hyperfine ground state in two $1064\unit{nm}$ crossed-beam optical dipole traps.
For technical reasons (atomic beam along $\uvect{z}$ and imaging beam along $\uvect{x}$) our dipole trapping beams are oriented at $\sim45^{\circ}$ to the horizontal coordinates $x$ and $z$.
The propagation directions of the two intersecting dipole trapping beams ($1/e^2$ radii of $75\unit{\upmu m}$ and $89\unit{\upmu m}$ respectively) define near-perpendicular horizontal axes $\uvect{x}'$ and $\uvect{z}'$ as shown in \reffig{fig:apparatus_full}.
The amplitude and horizontal position of each beam is controlled using a separate AOM.
Driving either one of the AOMs with two radiofrequency (rf) tones from an agile direct-digital synthesizer produces two diffracted orders, resulting in two crossed-beam dipole traps separated along either $\uvect{x}'$ or $\uvect{z}'$.
The rf frequencies determine the separation of the dipole traps~\cite{shin_atom_2004} along the intersecting beam.
\begin{figure}
    \centering
    \includegraphics[width=0.800\columnwidth]{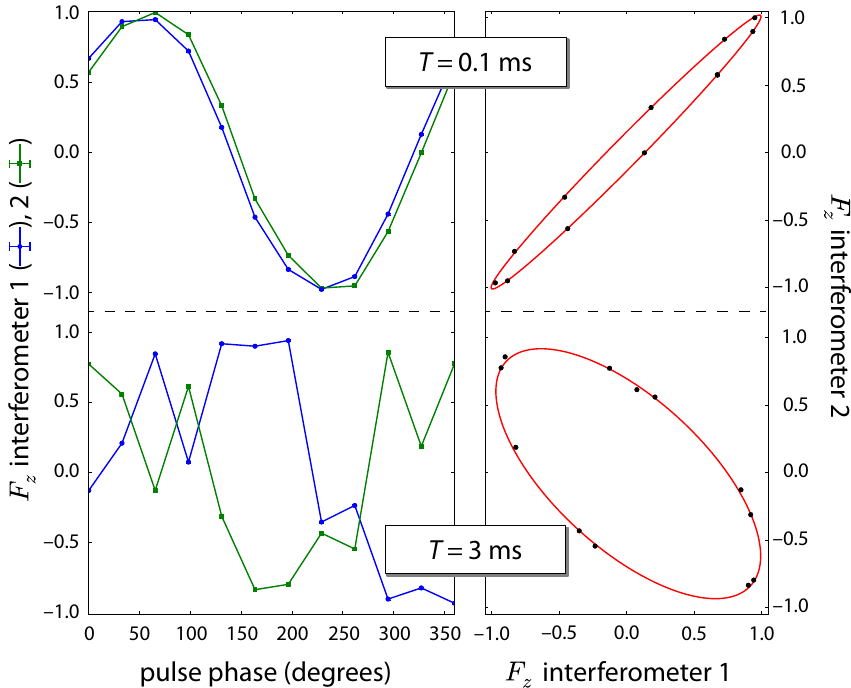}
    \caption{
        \label{fig:common_mode_rejection}
        (Color online)
        Spin projections at two different Ramsey times $T=0.1\unit{ms}$ and $3\unit{ms}$, plotted (left) as a function of $\pi/2$-pulse phase $\varphi$, and (right) parametrically from the output of each interferometer.
        If the phase acquired by each interferometer is stable (determined by the interrogation time $T$ and the shot-to-shot stability of $\vert \vect{B}(\vect{r}_i) \vert$), scanning $\varphi$ through $360^\circ$ maps one period of an interference fringe in $F_z$ (top left).
        Shot-to-shot fluctuations of $\vert \vect{B}(\vect{r}_i) \vert$ lead to an increasingly irreproducible phase acquired by each interferometer for longer interrogation times (bottom left).
        This onset of \emph{absolute} phase noise scrambles the phase domain fringes for $T > 1\unit{ms}$, but does not affect the the uncertainty in the \emph{relative} phase $\dphi$ calculated from fitting an ellipse, as the fluctuations in $\vert \vect{B}(\vect{r}_i) \vert$ are common-mode to each interferometer (right).
        Error bars correspond to the uncertainty in the measured spin projection in a single absorption image.
    }
\end{figure}

We Bose condense $5\times10^4$ atoms in each trap, initially separated by $100\unit{\upmu m}$ to maximize loading efficiency from a precursor hybrid optical dipole-magnetic quadrupole trap~\cite{lin_rapid_2009}.
The condensates are further separated over $2\unit{s}$ with a smooth frequency ramp, achieving a maximum separation of $\Delta x' = 680\unit{\upmu m}$ ($\Delta z' = 840\unit{\upmu m}$) when splitting the beam propagating along the $\uvect{z}'$ direction ($-\uvect{x}'$ direction).

Using three orthogonal coil pairs, we ensure the magnetic field is oriented along one of the $(x, y, z)$ axes with magnitude in the range $30$--$60\unit{\upmu T}$.
We extinguish the dipole trapping light and the two condensates begin to fall.
After $100\unit{\upmu s}$ freefall we initiate Ramsey interferometry between all Zeeman states $m = -1, 0, +1$ of the $F=1$ hyperfine ground state using a resonant rf $\pi/2$-pulse.
The two falling condensates comprise two independent interferometers.
The interferometers are closed with a second $\pi/2$-pulse after an interrogation time $T$.
Spin components ($m=0, \pm 1$) of each condensate are separated after further freefall by pulsing a $0.50\unit{T\,m^{-1}}$ magnetic field gradient for $3\unit{ms}$.
The number of atoms $N_{m,\alpha}$ in state $m$ of condensate $\alpha=1,2$ is determined by absorption imaging with a resonant laser after a total drop time of $23\unit{ms}$.
This constitutes a single realization, or \emph{shot}, of the experiment from which we compute the normalized spin projection $F_{z,\alpha} = \sum_{m} m N_{m,\alpha} / \sum_{m} N_{m,\alpha}$ for each interferometer.
\begin{figure}[tb!]
    \centering
    \includegraphics[width=0.800\columnwidth]{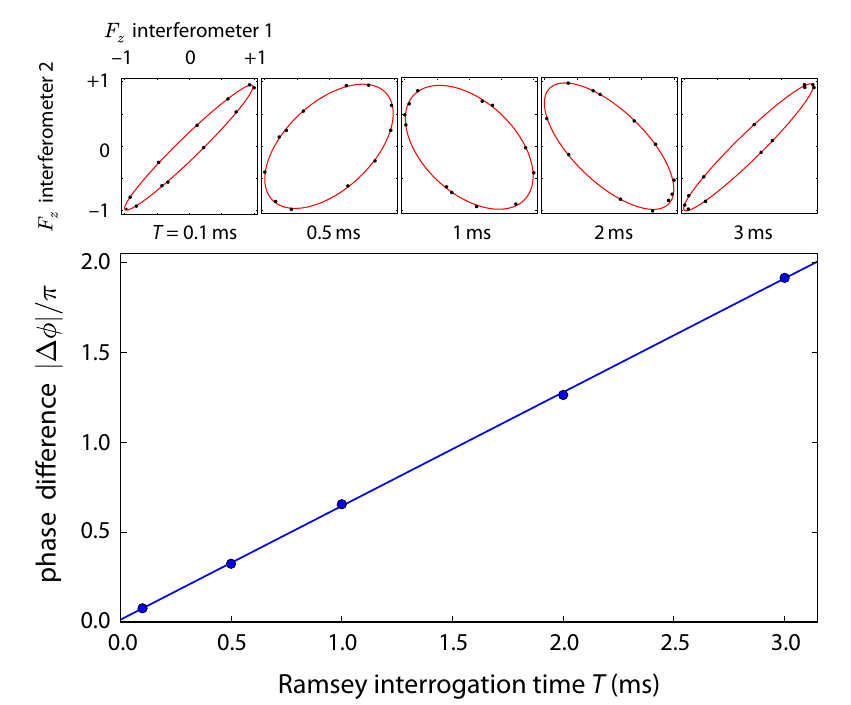}
    \caption{
        \label{fig:phase_vs_time}
        (Color online)
        Measurement of a magnetic field gradient using differential atom interferometry.
        Varying the phase of the second $\pi/2$-pulse of the Ramsey sequence for each fixed interrogation time $T$ traces an ellipse (top).
        From each ellipse we extract the magnitude of the phase difference $\abs{\dphi}$, and determine the field gradient using \refeq{eq:phase_to_gradient}; $\partial B_y/\partial z' = -53.3(3)\unit{nT\,mm^{-1}}$ for these data.
        Statistical uncertainties of $\abs{\dphi}$ and $F_{z,i}$ are smaller than the data points.
    }
\end{figure}

\section{Results}
\label{sec:results}
Ramsey fringes in the phase domain, $F_{z,1}(\varphi)$ and $F_{z,2}(\varphi)$, are clearly resolved for interrogation times $T < 500\unit{\upmu s}$, and are subsequently dominated by phase noise induced by magnetic field fluctuations common to both interferometers.
Plotting the interferometer outputs parametrically yields an ellipse, which is immune to common-mode phase noise.
This has been utilized in gravity gradiometry where atomic momentum states are interfered, and the common-mode phase noise derives from vibration of the reference platform~\cite{foster_method_2002, mcdonald_momentum_2013}.
To our knowledge this work is the first application of the elliptical data reduction to exclusively spin interferometry.
The general form of a conic is $aX^2+bXY+cY^2+dX+eY+f = 0$, with an ellipse satisfying $b^2-4ac <0$. The relative phase is given by~\cite{foster_method_2002}

\begin{equation}
\label{eq:dphi_ellipse}
    \cos \Delta \phi = \frac{b}{2\sqrt{ac}} \,.
\end{equation}

We fit an ellipse~\cite{fitzgibbon_direct_1999, szpak_guaranteed_2012} to a parametric dataset $(X, Y) = (F_{z,1}(\varphi), F_{z,2}(\varphi))$ to extract $\abs{\dphi}$; by repeating this process for different interrogation times we compute a magnetic field gradient via $d(\dphi)/dT$ in \refeq{eq:phase_to_gradient}.
We determined the sign of $\dphi$ from phase domain fringes at short interrogation times.
A measurement of a gradient in the $y$-component of the magnetic field is shown in \reffig{fig:phase_vs_time}.
We infer a nominal precision of $0.30\unit{nT\,mm^{-1}}$ from the combined statistical and systematic uncertainty in the slope of such linear fits.
Potential sources of systematic error are $T^3$ corrections to $\Delta \phi$ inherent to interrogation during freefall, and slow drifts in the gradient measurand.
We emphasize that measurements with varying interrogation times as shown here are performed to ensure the correct number of phase cycles are accounted for; in principle we may extract the field sensitivity from a single ellipse at the longest possible interrogation time.
This will be discussed in more detail in section \ref{sec:sensitivity}.

We quantified the common-mode rejection of the gradiometer by comparing the phase noise from a single interferometer to the uncertainty in the relative phase extracted from the elliptical fits.
Assuming our magnetic field noise is baseband, the measured phase noise yields a standard deviation of the Larmor frequency of $\sigma_{\omega_{\mathrm{L}}} = 2\pi \times 192(11)\unit{Hz}$, representing a common-mode rejection ratio exceeding $50\unit{dB}$.
\reffig{fig:common_mode_rejection} shows the deterioration of phase domain fringes at long interrogation times, while the corresponding parametric plots do not exhibit discernible degradation.

\begin{figure}[tb!]
    \centering
    \includegraphics[width=0.800\columnwidth]{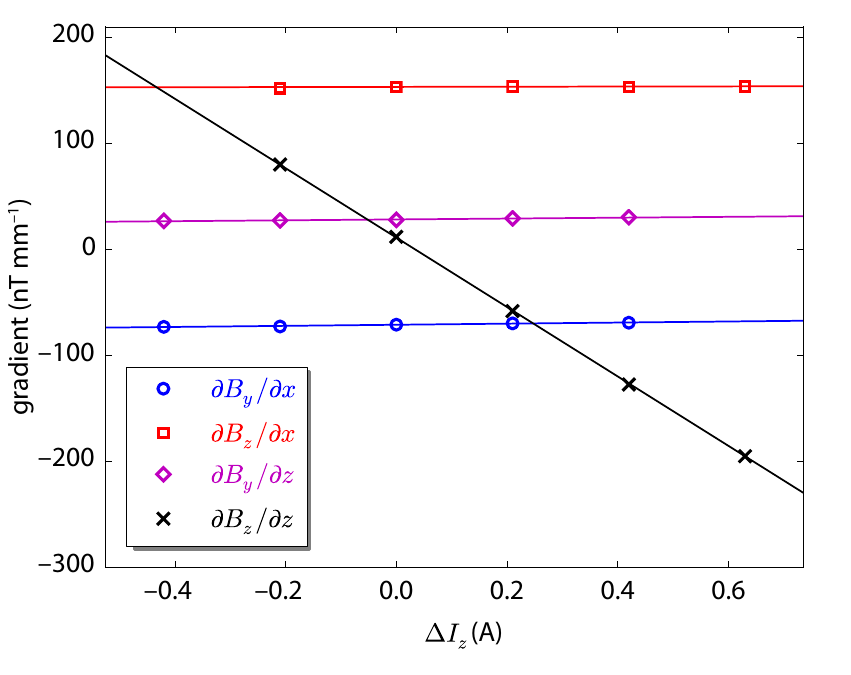}
    \caption{
        \label{fig:applied_gradients}
        (Color online)
        Response of the gradiometer to a gradient applied by driving a differential current $\Delta I_z$ through the $z$-bias coils.
        Four field derivatives $\partial B_y/\partial x$, $\partial B_z/\partial x$, $\partial B_y/\partial z$, and $\partial B_z/\partial z$ are measured by biasing the gradiometer along $y$ or $z$ with baselines along $x'$ or $z'$.
        The dominant gradient is \mbox{$\partial B_z/\partial z = -328(1) \Delta I_z \unit{nT\,mm^{-1}}$}; the relative insensitivity of the other measured gradients to $\Delta I_z$ quantifies the alignment of the bias coils along the Cartesian axes.
    }
\end{figure}

To demonstrate the tensor sensitivity of the gradiometer, we measured multiple field derivatives as a function of the current imbalance $\Delta I_z$ in the $z$-bias coils (\reffig{fig:applied_gradients}).
This allows us to explicitly evaluate the response of the gradiometer to an applied gradient when biased differently; with a field along $\uvect{z}$, we measure $\partial B_z/\partial z$ and $\partial B_z/\partial x$ proportional to and independent of $\Delta I_z$, respectively, in agreement with a numerical Biot-Savart calculation.
Orienting the magnetic field along the $y$-axis renders the gradiometer insensitive to the applied gradient, as this measures $\partial B_y/\partial x$ and $\partial B_y/\partial z$ (\refeq{eq:modB_approx}), which we find to be $< 5.1(6)\unit{nT\,mm^{-1}\,A^{-1}}$ due to imperfect alignment of the magnetic field along the $y$-axis.

The background magnetic field environment of our apparatus has predominantly linear magnetic field gradients originating from equipment within $1\unit{m}$ of the atoms.
The magnetic field gradient tensor can be represented by a matrix $G_{ij} = \partial B_i / \partial x_j$.
Using three bias field orientations and baselines that span the horizontal $x',z'$ plane is sufficient to calculate the full gradient tensor $G$ in the $x,y,z$ frame from Maxwell's laws for magnetic fields in a vacuum.
Our experimental conditions (in-vacuum, no significant electric fields) permit the use of $\nabla \times \mathbf{B} = 0$ and $\nabla \cdot \mathbf{B} = 0$ when applying Maxwell's equations.
We can thus determine gradients $\partial B_i/\partial y$ from the gradients measured in the horizontal plane, resulting in the gradient tensor
\begin{equation}
\label{eq:tensor_measured}
    G = \left(
            \begin{array}{rcr}
            -57.1(7)\: & \tikzmark{top}{-69.2(4)}\; & 147.0(7) \\
            -69.2(4)\: & 151.8(8) &  26.6(4) \\
            149.5(3)\: & \:\:\tikzmark{bottom}{26.6(4)} & -94.7(3) \\
            \end{array}
            \right) \unit{nT\,mm^{-1}} \, ,
    \submatrixborder[first]
\end{equation}
where the inferred values are inside the dashed box.
We observe that measurements of gradients over the course of a month do not vary beyond their uncertainties.
The gradient tensor has applications in localizing magnetic sources, which has seen widespread use in geophysics and surveying~\cite{stolz_magnetic_2006,schmidt_magnetic_2006,clark_new_2012}.
The power of the gradient tensor is exemplified by the ability to localize a source from a tensor measurement at a single point in space;
the three orthogonal eigenvectors of the matrix $G$ span a coordinate system in which off-diagonal gradient terms vanish, with the eigenvector corresponding to the largest magnitude eigenvalue pointing at the dominant dipole source~\cite{pedersen_gradient_1990}.
For the gradient tensor in \refeq{eq:tensor_measured}, this vector points in the direction of our dominant gradient source, the unshielded permanent magnets of an ion pump.
Homogeneous magnetic fields are required to preserve the coherence or robust entanglement of spins in many systems, e.g. the macroscopically entangled singlet state of a spinor quantum gas.
Here the knowledge of the gradient tensor is paramount: by applying diagonal gradients (i.e. those produced by two anti-Helmholtz coil pairs) along the eigenaxes of $G$, all components of the gradient tensor can be zeroed.
In many applications, it is sufficient to achieve a uniform field strength, i.e. to zero $\vert \nabla B \vert$.
Biasing the magnetic field along one of the eigenvectors of $G$ reduces this problem to canceling a single diagonal gradient.
Elsewhere, merely minimizing $\vert \nabla B \vert$ may be sufficient, and large reductions in $\vert \nabla B \vert$ can be achieved by changing the direction of the bias field alone, e.g. for the gradient tensor in \refeq{eq:tensor_measured} $\vert \nabla B \vert$ varies by a factor of 4 depending on the field direction.

Tensor gradiometers make use of Maxwell's laws to infer the complete tensor from as few as five independent gradient terms~\cite{fraser-smith_magnetic_1983, koch_room_1996, schmidt_getmag_2004, sui_compact_2014}.
Gradients measured over baselines in the $x$-$z$ plane quantify a systematic error in our inference; the equality of $\partial B_z/\partial x$ and $\partial B_x/\partial z$ is violated by $<2.5(7)\unit{nT\,mm^{-1}}$.
We attribute this discrepancy to asymmetric sampling of residual field curvature from the bias coils and imperfect cancellation of transverse field components when aligning $\vect{B}$ along a given axis (\refeq{eq:modB_approx}).
These two systematics may be reduced by larger Helmholtz coils, and by applying larger bias fields.
While an in-plane measurement is sufficient to determine the gradient tensor, two-axis acousto-optic deflection of the trapping beams could be used to measure all terms independently.

\section{Operational sensitivity as a co-magnetometer}
\label{sec:sensitivity}
Our scheme could also function as a precision \emph{co-magnetometer}, where one BEC placed in the vicinity of a small magnetic source acts as a sensor and the other condensate a reference interferometer, providing substantial common-mode rejection of ambient noise.
This is the mode of operation we have in mind when specifying field sensitivities per unit bandwidth or per unit spatiotemporal bandwidth below, the standard metrics for magnetometry.
No analagous metric is in use for \emph{gradient} sensitivity per spatiotemporal bandwidth.

The sensitivity of atomic magnetometers is determined by the ability to detect the Larmor precession of spins in a magnetic field.
For large atom numbers, as in warm vapor magnetometers, spin relaxation and photon shot noise ultimately limit sensitivity.
Cold atom systems interrogate smaller, trapped samples, and the spin projection noise at the standard quantum limit $\left(\delta F_z\right)_{ \text{SQL}} = 1/\sqrt{2N}$ for $N$ spin-1 atoms is relatively more important.
As multiple experimental shots are required to impute a differential phase from the elliptical data reduction, a single-shot phase sensitivity is ill-defined here.
Nonetheless, the differential phase uncertainty from fitting an ellipse with $M$ points scales with $1/\sqrt{M}$, and thus the quantum limited field sensitivity is $\delta B_{\text{SQL}} \sim 1/\left(\gamma \sqrt{N T D T_{\text{int}}}\right)$ for $N/2$ atoms per condensate, a total integration time $T_{\text{int}} = M T_{\text{shot}}$, a duty cycle $D = T/T_{\text{shot}}$, and a single-shot duration of $T_{\text{shot}}$.
From repeated absorption images we determine that the uncertainty in our measurements of spin projection $\delta F_z$ are $\sim 3$ times that of the standard quantum limit for $N=10^5$ atom condensates, and this ultimately determines the relative phase uncertainty extracted from ellipse fits.
The corresponding field sensitivity per unit bandwidth is $\delta B \sqrt{T_{\text{int}}} = 360\unit{pT\,Hz^{-1/2}}$ for $T_{\text{shot}}=25\unit{s}$ and $T = 3\,\text{ms}$.

\section{In-trap interferometry: prospective sensitivity}
\label{sec:intrap}
A gradiometer or co-magnetometer formed from a pair of \emph{trapped} condensates offers a smaller sensor volume and a significant improvement in sensitivity compared to one formed from atoms in freefall.
A prospective in-trap co-magnetometer can attain spatiotemporal sensitivities comparable with established forms of small-volume magnetic sensing, as we show in this Section.

For an in-trap magnetometer to be possible, it is necessary to suppress the vector light shift induced by residual elliptical polarization of the trapping beams.
These `fictitious' magnetic fields can be minimized by orienting the magnetic bias field to be perpendicular to the wavevector of both dipole beams, although this limits vector field sensitivity to only one spatial direction.
The vector light shift may be reduced by several orders of magnitude by ensuring the dipole trapping beams are near-linearly polarized \emph{in vacuo} with conventional \emph{ex vacuo} polarimetry.
The glass vacuum cell is inevitably birefringent, and superlative linearity of polarization at the atoms requires an atomic measurement~\cite{romalis_zeeman_1999, steffen_note:_2013}.
Differential Ramsey interferometry can provide such a measurement by sensing the phase difference between condensates exposed to trapping light of different intensities, but of common polarization.
Careful adjustment of a quarter-wave plate before the vacuum cell will be sufficient to achieve linear polarization at the atoms and thus a vanishing vector light shift.
This will be the focus of a forthcoming publication.

The sensitivity of each interferometer scales with the evolution time and the atomic density.
A Zeeman coherence time approaching one second was observed in a spin-1 \Rb condensate~\cite{higbie_direct_2005}, limited by losses due to density-dependent collisions.
For a \Rb BEC with a peak number density of $10^{14}\unit{atoms\,cm^{-3}}$ (corresponding to $N=10^6$ atoms for the current trap), an interrogation time of $T=200\unit{ms}$ is foreseeable, an order of magnitude lower than the three-body limited lifetime.
The prospective in-trap magnetometer could thus achieve a differential field sensitivity per unit bandwidth of $600\unit{fT\,Hz^{-1/2}}$ at the standard quantum limit, even with the same trap and single-shot duration used here (corresponding to a non-unity duty cycle of $D=0.008$).

Spatial resolution is conventionally quantified by the sensing volume $V$; a vapor magnetometer with $V=300\unit{mm^3}$ attained sub-femtotesla sensitivities~\cite{kominis_subfemtotesla_2003} whereas NV-center magnetic probes deliver nanoscopic resolution but at much lower field sensitivities~\cite{taylor_high-sensitivity_2008}.
For the freefall measurements described herein, the sensor volume $V = 2\times10^{-5}\unit{mm^3}$ is that swept out by a falling, expanding condensate during the Ramsey interrogation.
Using the metric of spatiotemporal sensitivity, our demonstrated measurement has $\delta B \sqrt{T_{\text{int}}} \sqrt{V} = 51\unit{fT\,cm^{3/2}\,Hz^{-1/2}}$.
The above prospective in-trap magnetometer has $V = (20\unit{\upmu m})^3$, corresponding to $\delta B \sqrt{T_{\text{int}}} \sqrt{V} = 0.05\unit{fT\,cm^{3/2}\,Hz^{-1/2}}$.
This large prospective improvement is not unprecedented in microscale magnetometry where first demonstrations are far from fundamental limits; warm vapor magnetometry in microfabricated cells ($V\sim\unit{mm^3}$) was first demonstrated at a sensitivity of $5500\unit{fT\,cm^{3/2}\,Hz^{-1/2}}$~\cite{schwindt_chip-scale_2004}, rapidly developed to $5.0\unit{fT\,cm^{-3/2}\,Hz^{-1/2}}$~\cite{shah_subpicotesla_2007}, and more recently $0.16\unit{fT\,cm^{3/2}\,Hz^{-1/2}}$~\cite{griffith_femtotesla_2010}.
The spatiotemporal sensitivity can also be expressed in units of energy per unit bandwidth $\epsilon = (\delta B)^2 T \, V/2 \mu_0$.
While the low duty cycle and number of spins in BEC based measurements limits their field sensitivity per unit bandwidth, the far smaller volume results in a $\epsilon$ comparable to warm vapor magnetometers, with $\epsilon \sim 50$--$100\hbar$ for vapor magnetometers~\cite{griffith_femtotesla_2010,dang_ultrahigh_2010}, and $\epsilon \sim 10 \hbar$ for the prospective magnetometer described above.

During the preparation of this manuscript, we became aware of related work also employing an array of condensates simultaneously addressed with a Ramsey sequence~\cite{muessel_scalable_2014}. These authors focused on the application of spin-squeezing to realize phase sensitivities below the atomic shot noise (at short interrogation times), in contrast to our work which focuses on dynamic reconfigurability of two condensates to achieve tensor sensitivity of magnetic field gradients.

\section{Conclusions}
\label{sec:conclusions}
We have demonstrated magnetic tensor gradiometry using differential Ramsey interferometry of spatially separated BECs in freefall.
The gradiometer senses vector components of the magnetic field, rejecting gradient components orthogonal to the biasing direction.
The gradiometer is immune to common-mode magnetic noise orders of magnitude larger than the field difference, and operates without field cancellation or screening.
The dynamic reorientability of the gradiometer baseline with microscale resolution allows for precision surveys of magnetic microstructures and the ambient magnetic environment of trapped quantum gases.
The gradiometer could be used as a high-precision co-magnetometer with substantial common-mode rejection, allowing for microscale magnetic sensing \emph{in vacuo}.

\begin{acknowledgments}
This work was supported by the Australian Research Council (DP1094399) and the Australian Postgraduate Award Scheme. We are grateful to Z.~Szpak and W.~Chojnacki for assistance with ellipse fitting, Y.~Levin for useful conversations, and K.~Helmerson for a careful reading of the manuscript.
\end{acknowledgments}

\bibliography{gradiometer}

\end{document}